\documentclass[aps,prl,twocolumn,showpacs,superscriptaddress]{revtex4}
\usepackage{graphicx}
\usepackage{amsmath}
\newlength\figurewidth
\begin{document}

\title{ The interplay between real and pseudo magnetic field in graphene with strain   }

\author{Kyung-Joong Kim}
\affiliation{Department of Physics, Chungnam National University,
 Daejeon 305-764, Republic of Korea}
\author{Ya. M. Blanter}
\affiliation{Kavli Institute of Nanoscience, Delft University of
Technology, Lorentzweg 1, 2628 CJ Delft, The Netherlands}
\author{Kang-Hun Ahn  $^{1,}$}%
\email[\* Corresponding author. Email:\ ]{ahnkh@cnu.ac.kr}%
\affiliation{Department of Physics, University of Bath, Bath BA2
7AY, United Kingdom}
\date{\today}

\begin{abstract}
We investigate electric and magnetic properties of graphene with
rotationally symmetric strain. The strain generates large pseudo
magnetic field with alternating sign in space, which forms strongly
confined quantum dot connected to six chiral channels. The orbital
magnetism, degeneracy, and channel opening can be understood from
the interplay between the real and pseudo magnetic field which have
different parities under time reversal and mirror reflection.
 While the orbital
magnetic response of the confined state is diamagnetic, it can be
paramagnetic if there is an accidental degeneracy with opposite
mirror reflection parity.

\end{abstract}

\maketitle

The recent successful preparation of one-atom layer of carbons,
graphene \cite{novoselov,zhang}, has provided the opportunity of
theoretical and experimental research of relativistic physics in
nanoelectronics. While  quantum dots  which confine quasi-particles
in graphene are basic building blocks for its nanoelectronic
application, the confinement turns out to be nontrivial.
It is because, in
graphene where the quasi-particles are described by massless Dirac
fermions, they can penetrate large and wide electrostatic barriers
due to the effect of Klein tunneling\cite{klein}.
 In principles, graphene dots can be realized
by a spatially inhomogeneous magnetic
field\cite{martino,giavaras,dwang}, but the required magnetic field
for the confinement, however, is unreasonably strong compared to
usual electronics application. Recently, strain engineering of
graphene\cite{pereira09,levy} attracted great attention as an
alternative tool for graphene electronics because the strain induces
strong {\it pseudo} magnetic field which guides electrons. Thus, for
 the strained graphene to work successfully in combination with existing technologies, it
is now important to understand the physical properties of the pseudo
magnetic field.

In this Letter, we investigate the relative contribution of real and
pseudo magnetic field to the electric and magnetic properties of the
graphene. We show that reasonable size of strain can generate strong
pseudo magnetic field to form a graphene quantum dot with six chiral
channels.
 It will be  demonstrated that the
different symmetry of real and pseudo magnetic fields give rises to
rich properties of channel opening and orbital magnetism.
 The pseudo magnetic field
appears since the variation of hopping energies by elastic strains
enters the Dirac equation
\cite{guinea08,eakim,manes,guinea10,doussal,Morpurgo}.
 While the strong confinement is due to the
fact that the pseudo-magnetic field is very strong ($\sim 10 T$),
the six chiral channels are due to the topology of the pseudo
magnetic field, where charged particles propagate along the
zero-field line. As we will show here, the real and pseudo magnetic
field have different parities under the symmetry operation such as
time reversal and mirror reflection. From the symmetry arguments, we
prove that while the real magnetic field breaks the time reversal
symmetry in its Hamiltonian, it does not lift the valley degeneracy.
 We will
demonstrate our theory by showing orbital diamagnetism of the
confined state. It will be shown that paramagnetic response is also
allowed when an partially open state with opposite parity becomes
degenerate with the confined state.

\begin{figure}
\includegraphics[width=\figurewidth]{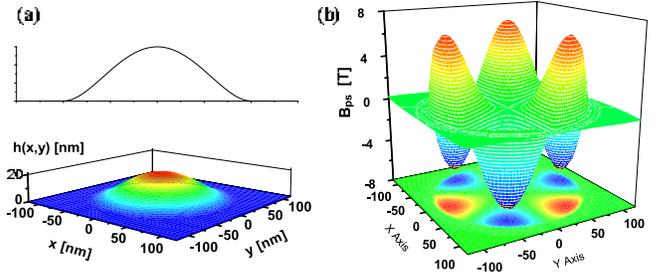}
\caption{The vertical displacement of our considered systems in the
text (a) and the pseudo magnetic fields (b). The radius is $R$=100nm
and the height is $h_{0}$=20nm. }
 \label{effB}
\end{figure}

 Let us consider graphene where mechanical deformation is allowed
in a restricted disk shape. This can be realized by a circular hole
made in  substrate below the graphene sheet, and the deformation is
induced through external force.  In experiments, circularly
symmetric strain fields can be applied by an AFM tip or  by a
homogeneous gas pressure acting on graphene below the
substrate\cite{Bunch}. When strain is induced by homogeneous load,
the optimized vertical displacement $h({\bf r})$is given
by\cite{landau}
\begin{eqnarray}
h({\bf r})&=&\frac{f_{0}}{4^{3}D}\left( R^{2}-(x^2 +
y^2)\right)^{2} \ ,
\end{eqnarray}
where $f_0$ is the force per unit area acting on surface, $D$ is the
bending rigidity, $h_{0} = f_0 R^4/(4^3D)$ is the vertical
displacement at center, and $R$ is the radius of the region where
the deformation is allowed. The in-plane relaxation of the carbon
atoms  $u_x, u_y$ can be calculated by minimizing the elastic free
energy for the given vertical displacement h({\bf r}),
\begin{eqnarray}
\textit{F} = \int dxdy \left[\frac{\kappa}{2}(\nabla^{2}h)^2 +
\frac{\lambda}{2}(\sum_{i} u_{ii})^2 + \mu\sum_{ij}(u_{ij})^2 \right]
\end{eqnarray}
where $\kappa$ is the bending rigidity, $\lambda$ and $\mu(\approx
3\lambda)$ are Lam$\acute{e}$ coefficients, and $u_{ij}$ is the
strain tensor. Here, the strain tensor $u_{ij}({\bf r})$ is related
to the displacement fields via $u_{xx}=\partial u_{x}/\partial x
+\frac{1}{2}(\partial h/\partial x)^{2}$, $u_{yy}=\partial
u_{y}/\partial y +\frac{1}{2}(\partial h/\partial y)^{2}$, and
$u_{xy}=\frac{1}{2}(\partial u_{x}/\partial y +\partial
u_{y}/\partial x)+\frac{1}{2}(\partial h/\partial x)(\partial
h/\partial y)$.

\begin{figure}
\includegraphics[width=\figurewidth]{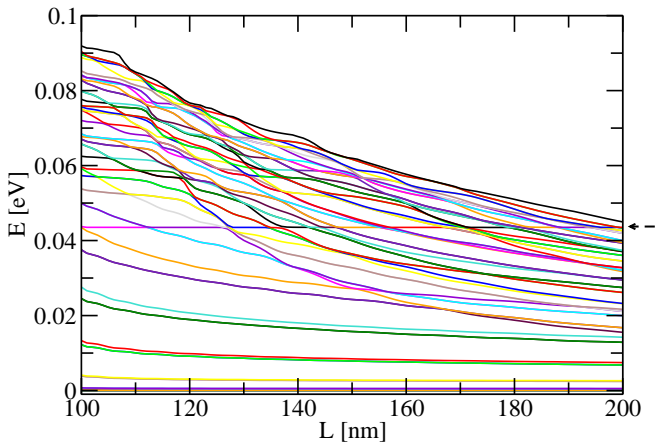}
\caption{The eigenenergies as a function of the system size $L$. We
set $R$=100nm and $h_{0}$=20nm. Note that at certain energy (marked
by an arrow), the eigenenergy becomes insensitive to the system size
$L$ .} \label{energies}
\end{figure}

We consider spinless fermions in a graphene lattice. The spinless
quasi-particle in the graphene can be described by 4-component
wavefunction ${\bf \Psi^{\rm
T}}=(\Psi_{K}^{T},\Psi_{-K}^{T})=(\psi_{A,K},\psi_{B,K},\psi_{A,-K},\psi_{B,-K})$.
These are the electron wavefunctions near two inequivalent points
$\pm K$ in hexagonal Brillouin zone in the two crystalline
sublattices A and B. We ignore the valley mixing by the strain based
on the results of a tight-binding calculation showing that the two
inequivalent valleys are not coupled under uniaxial deformations up
to 20\% \cite{pereira}.
 The main effect of the strain field on the electrons is to modify the energy for electron hopping between the nearest
neighbor atoms. The modified quasi-particle energy by the strain is
well described by introducing the pseudo gauge field
 ${\bf A}_{\rm ps}({\bf r})$ in the  massless Dirac
equation,

\begin{eqnarray}
H=v_{F}
\begin{pmatrix}
{\vec{\sigma}}\cdot ({\bf p}+e {\bf A}_{\rm ps}  ) & 0 \\ 0 &
-{\vec{\sigma}}^{*}\cdot ({\bf p} -e {\bf A}_{\rm ps})
\end{pmatrix}
\label{diracweyl}
\end{eqnarray}
where $-e$ is the electric charge, $v_{F}\approx 9 \times 10^{5}$m/s
is the Fermi velocity, ${\bf
 p}=-i\hbar (\partial_{x},\partial_{y})$, and
 $\vec{\sigma}=(\sigma_{x},\sigma_{y})$ are Pauli matrices acting in
 the sublattice space. Here, we take two K points in
 opposite direction along x-axis.
The pseudo gauge field ${\bf A}_{\rm ps}({\bf r})$ is written as
\begin{eqnarray}
{\bf A}_{{\rm ps}}=\frac{t\beta}{ev_{F}} (u_{xx}-u_{yy},-2u_{xy}),
\label{gauge}
\end{eqnarray}
where $t\approx 2.8$eV is the electron hopping energy between the
nearest $\pi$ orbitals, and $\beta \approx 2-3$ is the dimensionless
coupling parameter for the lattice
deformation\cite{ando,eakim,manes}. In Fig.\ref{effB}, we plot the
pseudo magnetic field ${\bf B}_{\rm ps}=B_{\rm ps}{\bf k}=\nabla
\times {\bf A}_{\rm ps}$. This is the pseudo magnetic field
experienced by the particle in $K$ valley, and for the particle in
$-K$ valley, the sign of the magnetic field is opposite.

To investigate the confinement of the quasi-particles in graphene,
we compute the eigenenergies of the system as the total graphene
size used in the calculation $L$ increases and observe whether the
energy is insensitive to the system size. The graphene in the
calculation is a disc defined as $r < L$ and the graphene is
deformed only in the region of $r< R$. We impose the boundary
condition $\Psi_{A}(r=L)=\Psi_{B}(r=L)=0$. While the wavefunctions
which extend over the total system are sensitive to the boundary
conditions, localized states in the deformed are not affected by the
boundary condition.
 We obtain the eigenvalues of Hamiltonian using
basis functions $\psi_{A,B}=\sum c^{A,B}_{n,l}\phi_{n,l}({\bf r})$,
where $ \phi_{n,l}({\bf
r})=J_{l}(\frac{\alpha_{nl}}{L}r)e^{\textit{il}\theta}$. Here,
$\alpha_{nl}$ is the n-th zero of the Bessel function $J_{l}(x)$ of
order $l$, and we use the indices $l=0,\pm 1, \pm 2, \cdots$,
$n=1,2,3,\cdots$.

 As shown in Fig.2, we find that at certain energies, there exist eigenstates of which
eigenenergies are insensitive to the system size $L$. These are the
localized states induced by the deformation of the graphene.
Compared to the mid-gap states in a ripple array studied in
Ref.\cite{guinea08} which shows weak size dependence in logarithmic
scale, the eigenenergies of the localized states show almost no
size-dependence. From an analytic calculation, the length and energy
scales for the localized state can be obtained by bringing the
asymptotics of differential equation for the wavefunction to
dimensionless form. This is done by rescaling $r \to r_{0}
\tilde{r}$, $E \to E_0 \tilde{E}$, with
\begin{equation} \label{scales}
r_{0} = \left( \frac{\hbar v_F R^4}{8t\beta h_0^2} \right)^{1/3}, \
\ \ E_0 = \hbar v_F/r_{0} \ .
\end{equation}
The scale $r_{0}$ thus plays a role of the localization length and
can be estimated as $(R^4 a/h_0^2)^{1/3}$, where $a$ is the lattice
constant. Since the localization length must be shorter than the
hole radius, otherwise the Dirac equation with effective magnetic
field can not apply, the localization length expressed in the above
equation is valid only for strong enough load, $h_0 \gg \sqrt{R a}$.
The scale $E_0$ is associated with the depth of the potential well
and the energy of the localized energy levels.

It proves useful to consider the symmetries to understand the energy
spectra of quantum systems.
 The time reversal symmetry is not broken by the strain if we consider the problem with both of the valleys.
 A time reversal operation defined by
\begin{eqnarray}
 \mathcal{T} = \tau_{x}\mathcal{K}
 \label{treversal}
\end{eqnarray}
 satisfies
$\label{sym1} \mathcal{T} H \mathcal{T}^{-1} = H, ~~ \mathcal{T} i
\mathcal{T}^{-1} = -i $\cite{bjorken}.

 Here $\mathcal{K}$ is the complex conjugate
operator and $(\tau_{x},\tau_{y},\tau_{z})$ are the Pauli matrices
acting in the the valley space. Note that, the Kramers degeneracy is
not relevant here since the original system  has orthogonal symmetry
$\mathcal{T}^{2}=1$.

The quasi-particle can stay in a given valley provided there is no
short range scattering ( e.g. lattice defects). In this restricted
case, the single valley time reversal symmetry is broken by the
pseudo magnetic field. The single valley time reversal operator for
1/2 spin $ \mathcal{S}= - i \sigma_{y} \mathcal{K} $ does not
commute with the Hamiltonian in Eq.(\ref{diracweyl}). The
Hamiltonian is symmetric under the symplectic time reversal
transformation only in the absence of the strain
($\mathcal{S}^{2}=-1$); $ \mathcal{S}H({\bf A_{\rm
ps}}=0)\mathcal{S}^{-1}= -H({\bf A_{\rm ps}}=0) $. The relevant
Kramers degeneracy here is lifted by the pseudo magnetic field ${\bf
A_{\rm ps}}\neq 0$.

The Hamiltonian in Eq.(\ref{diracweyl}) is also symmetric under a
mirror reflection
\begin{eqnarray}
  \mathcal{M} = \sigma_{x}
\pi_{x},\\
\mathcal{M} H \mathcal{M}^{-1}= H, \label{Msymmetry}
\end{eqnarray}
where $\pi_{x}$ acts as $x\rightarrow x,~~y\rightarrow -y$. The
valley index remains same under the mirror reflection but inevitably
changes lattice index. The spatial symmetry
($|\psi_{A,K}(x,y)|^{2}=|\psi_{B,K}(x,-y)|^{2}$) of the probability
density of the localized state shown in Fig. 3 (a) and (b) reflects
the $\mathcal{M}$-symmetry in the Hamiltonian.

To understand the electronic structure of the strained graphene, we
rewrite the Hamiltonian in Eq.(\ref{diracweyl}) for a given valley
$K$ and eigenenegy $E$,
\begin{eqnarray}
\left[ v_{F}^{2}\left( {\bf p}+e{\bf A_{\rm ps}} \right)^{2} +\hbar
e v_{F}^{2} (\nabla \times {\bf A_{\rm ps}})_{z} \sigma_{z}
\right]\Psi_{K} = E^{2} \Psi_{K}. \label{reHamilt}
\end{eqnarray}
The first term in the left side of Eq.(\ref{reHamilt}) comes from
the kinetic energy and the second term is due to the pseudo Zeeman
coupling. The pseudo magnetic field is strongest at six points
forming a hexagon (Fig. 1) where local Landau levels might be
formed.
 For a given pseudo spin and valley, one can see maximum probability density around only
three points (Fig. 3(a) and (b)). This is because of a necessary
condition for the stable confinement on the pseudo Zeeman coupling;
\begin{eqnarray}
(\nabla \times {\bf A_{\rm ps}})_{z} \sigma_{z} < 0.
\end{eqnarray}
In this case, it costs higher energy when a quasi particle to go out
to weaker field region.
 The triangular (instead of hexagonal) shape of the
wavefunction for a given lattice and valley is due to the selective
stabilization by the pseudo Zeeman coupling.

\begin{figure}
\includegraphics[width=\figurewidth]{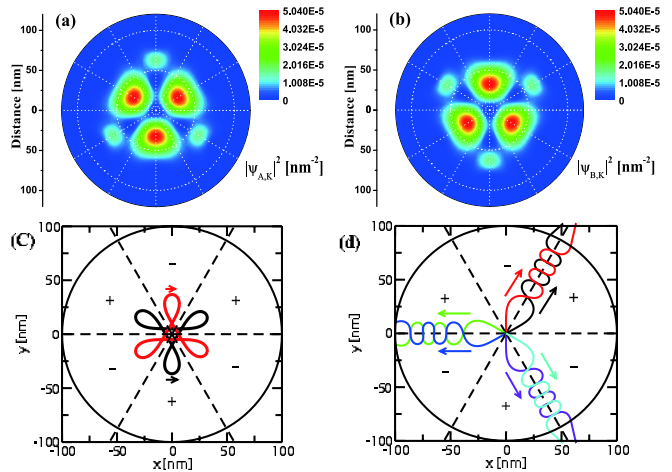}
\caption{  (a) Probability densities of the localized wavefunctions
$|\psi_{A,K}({\bf r})|^{2}$ and (b)$|\psi_{B,K}({\bf r})|^{2}$. We
set $R$=100nm, $h_{0}$=20nm. (c) Classical closed orbits of a
charged particle in the inhomogeneous magnetic field given by
$B_{\rm ps}$ which resemble the probability density of the localized
states shown in (a) and (b). (d) Classical paths describing three
outgoing channels.}
 \label{density1}
\end{figure}

The confinement of channeling in the strained graphene can be
visualized by investigating classical trajectories of the charged
particles in the pseudo magnetic field (Fig. 3 (c) and (d)). Among
the periodic orbits around the pseudo magnetic field maxima, we find
 {\it clover}-shaped orbits which resemble the localized
wavefunctions in (Fig. 3 (c)). These closed orbits are very unstable
against small perturbation. In quantum mechanics, the clover-shape
motion might be responsible for quantum transition  between the
sites of the local density maxima. We also find out-going
trajectories (Fig. 3 (d)) for different initial velocities. A
charged particle can propagate along the line where magnetic field
changes sign, which is, so called, {\it snake orbit}\cite{sim}. Due
to the symmetry of the pseudo magnetic field, there are incoming
trajectories in 60 degree rotated angle from those of outgoing
trajectories. Quantum mechanically, for given components of lattice
and valley, the graphene quantum dot is connected to three incoming
and three outgoing chiral channels.

The opening of the channels manifest in the energy spectra. The
localized energy level undergoes crossing and avoided crossing as
the graphene size $L$ changes. ( see Fig. 2 and Fig. 4 (a) for mode
detail.) When the channels open, the eigenenergy of the confined
state is affected by the graphene size, so has avoided crossing.
Meanwhile when the two level cross each other, the channels remain
closed and the eigenenergy of the localized state is insensitive
graphene size. In the process of avoided crossing, the confined
state undergoes transition to an outer state and a new outer state
becomes localized. Since any parity does not change in this
continuous process, the confined state transit only to the state
with same parity.

Let us consider the response of the strained graphene to real
magnetic field. When real magnetic field, ${\bf B}_{\rm
re}=\nabla\times {\bf A}_{\rm re}$, is applied, the minimal coupling
of the electromagnetic gauge field ${\bf A}_{\rm re}$  is done by
replacing ${\bf p}$ with ${\bf p}+e{\bf A}_{\rm real}$ in
Eq.(\ref{diracweyl}). The Hamiltonian becomes $H+H^{\prime}$ where
\begin{eqnarray}
H^{\prime}=v_{F}e
\begin{pmatrix}
{\vec{\sigma}}   & 0 \\ 0 & -{\vec{\sigma}}^{*}
\end{pmatrix}
\cdot {\bf A_{\rm re}},
\end{eqnarray}
and ${\bf A_{\rm re}}=\frac{1}{2}(-y,x)B$ is the gauge field for the
real magnetic field in z-direction. We want to address here that the
application of real magnetic field breaks the time reversal symmetry
of the Hamiltonian, but it does not lift the valley degeneracy. This
can be proved by showing $<\Psi_{\pm K}|\frac{\partial
H^{\prime}}{\partial B}|\Psi_{\pm K}> = 0$. The proof comes from the
fact the eigenstates have either even or odd parity of the mirror
reflection symmetry in Eq.(\ref{Msymmetry}), $\mathcal{M}^{2}=1$:
\begin{eqnarray}
\nonumber&& <\Psi_{ K}|\frac{\partial H^{\prime}}{\partial
B}|\Psi_{K}>  = <\Psi_{ K}| (-y\sigma_{x}+x\sigma_{y})|\Psi_{ K}>
\\ \nonumber &=&<\Psi_{
K}|\mathcal{M} (+y\sigma_{x}-x\sigma_{y})\mathcal{M}|\Psi_{ K}> \\
 &=&-<\Psi_{ K}| (-y\sigma_{x}+x\sigma_{y})|\Psi_{ K}>.
 \label{0proof}
\end{eqnarray}
Since $<\Psi_{ K}|\frac{\partial H^{\prime}}{\partial B}|\Psi_{K}>$
is equal to its own minus value, it must be zero. The leading
magnetic field dependence of the eigenenergy in the presence of
magnetic field is not linear but quadratic, $\propto B^{2}$. It
comes from the kinetic energy and its sign is positive. Therefore,
the orbital magnetization of the strain-induced quantum dot at zero
temperature, is diamagnetic ( $-\partial E/\partial B <0$ ) and
proportional to the applied magnetic field strength (See Fig. 4
(b)).

\begin{figure}
\includegraphics[width=\figurewidth]{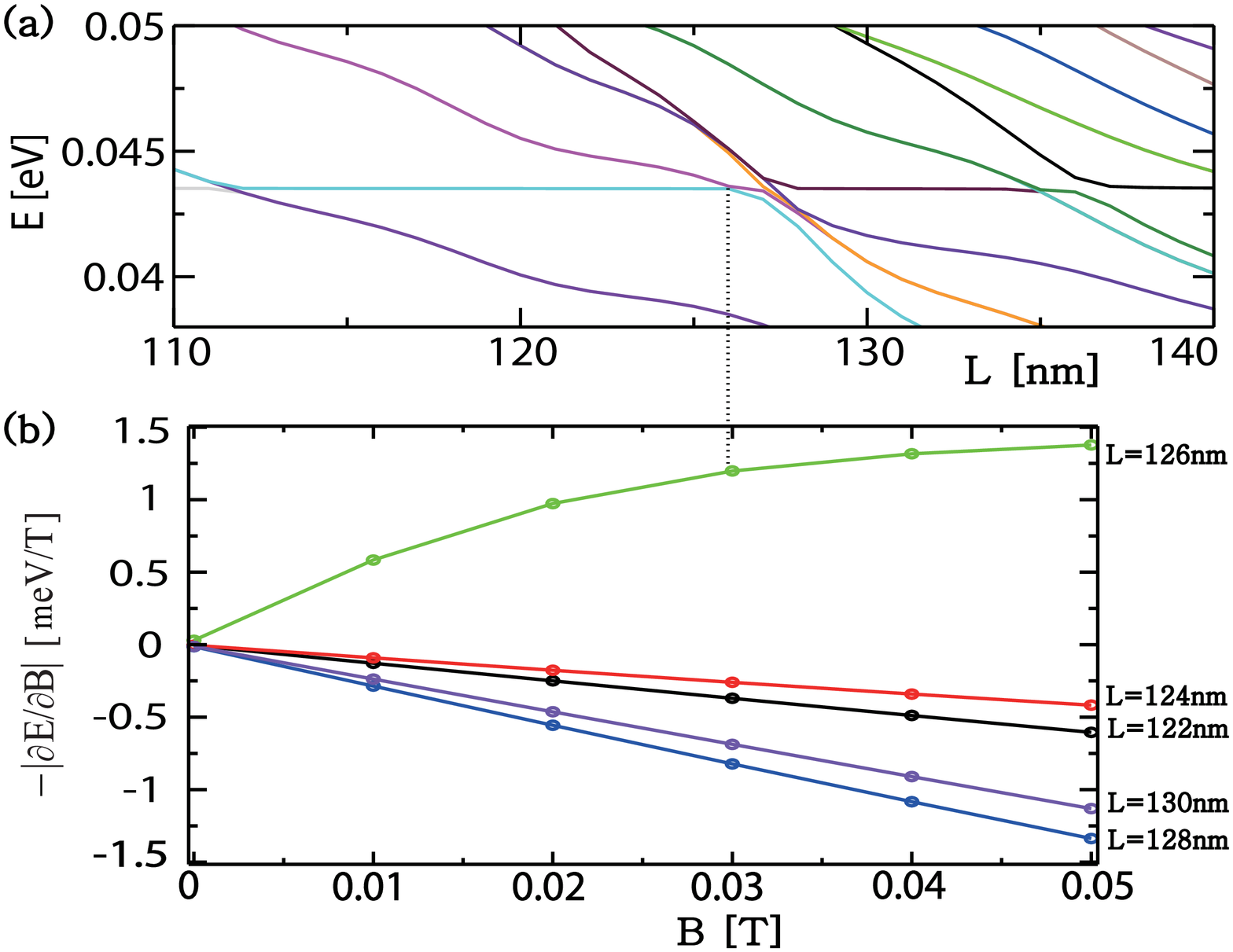}
\caption{ (a) A detailed energy spectra as a function of the system
size $L$ which shows crossing and avoided-crossing. (b) The orbital
magnetic response, $-\frac{\partial E}{\partial B}$ , of localized
states.  Mostly the magnetic response is diamagnetic
$-\frac{\partial E}{\partial B}<0$ but the response for a level
crossing point (indicated by dotted line) shows paramagnetic
response (positive value). We set $R$=100nm and $h_{0}$=20nm.}
\label{fig4}
\end{figure}

In contrast to the diamagnetic response of the confined state,
 the orbital magnetic response can be paramagnetic ( $-\partial
E/\partial B > 0$ ) when there are level-crossings. Near the region
of the level crossings, there are two energy levels with {\it
opposite} parity of $\mathcal{M}$.
 One of the states is a localized state and
the other is a partially opened state (not shown).   The accidental
degeneracy which happened here can be lifted by applying real
magnetic field because $H^{\prime}$ is odd under the mirror
reflection, $\mathcal{M}H^{\prime}\mathcal{M}=-H^{\prime}$. Then
energy splitting arises, proportional to the real magnetic field
strength, which contributes to paramagnetic response.

In conclusion, we have shown that the rotationally symmetric strain
in graphene can be considered a quantum dot with spatially separated
six chiral channels. The chiral channels exist along the line where
the pseudo magnetic field changes sign. The real and pseudo magnetic
field has different symmetry under a mirror reflection, which makes
the orbital magnetism be diamagnetic or paramagnetic depending on
the degeneracy.
 The orbital magnetic response of the
confined state is diamagnetic due to its kinetic energy. When there
is an degeneracy with opposite mirror reflection parity, the orbital
magnetism can be paramagnetic.

Quite recently, we became aware of a work on dynamics of electrons
in strain-induced pseudo magnetic fields\cite{Blaauboer}.

The authors thank M. Sieber, K. Richter, Y. Son, F. Guinea and M.
Fogler for useful discussions. This work was supported by the
National Research Foundation funded by the Korea government
(No.KRF-2008-C00140).



\begin{thebibliography}{100}
\bibitem{novoselov} K.S. Novoselov et. al., Science {\bf 306}, 666
(2004);Nature(London) {\bf 438}, 197 (2005).
\bibitem{zhang} Y. Zhang, et. al., Nature(London) {\bf 438}, 201 (2005).
\bibitem{klein} V. V. Cheianov and V. I. Falko, Phys. Rev. B {\bf 74}, 041403 (2006);
M. I. Katsnelson, K. S. Novoselov and A. K. Geim, Nature physics
{\bf 2}, 620 (2006)
\bibitem{martino} A. De Martino, L. Dell'Anna, and R. Egger, Phs.
Rev. Lett. {\bf 98}, 066802 (2007).
\bibitem{giavaras} G.Giavaras, P. A. Maksym and M. Roy, J. Phys.
Condes. Matter {\bf 21} 102201 (2009)
\bibitem{dwang} D. Wang, G. Jin, Phys. Lett. A, {\bf 373}, 4082, (2009).
\bibitem{pereira09} Vitor M. Pereira, A.H. Castro Neto, Phys. Rev.
Lett. {\bf 103}, 046801 (2009).
\bibitem{levy}N. Levy et.al., science, {\bf 329}, 5991 (2010).
\bibitem{guinea08} F. Guinea, M. I. Katsnelson, M. A. H. Vozmediano, Phys. Rev. B {\bf 77}, 075422 (2008).
\bibitem{eakim} Eun-Ah Kim and A. H. Castro Neto, Europhys. Lett., {\bf 84},
57007, (2008)
\bibitem{manes} J.L.Man\~es, Phys. Rev. B, {\bf 76}, 045430 (2007).
\bibitem{doussal} F. Guinea, B. Horovitz, P. Le Doussal, Phys. Rev. B {\bf 77}, 205421 (2008).
\bibitem{Morpurgo} A. F. Morpurgo and F. Guinea, Phys. Rev.
Lett., {\bf 97}, 196804 (2006).
\bibitem{guinea10} F. Guinea, M.I. Katsnelson, and A. K. Geim,
Nature Phys., {\bf 6}, 30 (2010).
\bibitem{Bunch} J. S. Bunch et al., Nano Letters {\bf 8}, 2458, (2008).
\bibitem{landau} L.D. Landau and E.M. Lifshitz, {\it Theory of Elasticity},
(Pergamon Press Ltd, London, 1959).
\bibitem{pereira} V. M. Pereira and A. H. Castro Neto, Phys. Rev. B {\bf 80}, 045401 (2009).
\bibitem{ando} H. Suzuura and T. Ando, Phys. Rev. B {\bf 65}, 235412
(2002); T. Ando, J. Phys. Soc. Jpn, {\bf 75}, 124701 (2006).
\bibitem{bjorken}  J. D. Bjorken and S. D. Drell, {\it Relativistic Quantum Mechanics}, McGraw-Hill, (1964), p. 72.
\bibitem{sim} H.-S. Sim, K.-H. Ahn, K. J. Chang, G. Ihm, N. Kim, and S. J. Lee,
Phys. Rev. Lett. 80, 1501 (1998); F. Evers et. al., Phys. Rev. B,
{\bf 60}, 8951 (1999).
\bibitem{Blaauboer} G. M. M. Wakker, R. P. Tiwari, and M. Blaauboer,
arXiv:1105.3588.
\end{thebibliography}



\end{document}